\documentclass[12pt]{article}
\usepackage{graphicx}
\usepackage{epsfig}

\newcommand{\pmu}{{P_{\mu}}}

\title{%
\vspace*{-2cm}
\begin{flushright}
{\small KYUSHU-HET 57}\\\vspace{-0.5cm}
{\small OUHEP-01-1}\\
\end{flushright}
{~}\\
 Search for T-violation in Neutrino Oscillation with the Use of
 Muon Polarization at a Neutrino Factory
}

\author{%
   Toshihiko Ota%
     \footnote{e-mail address: {\tt 
toshi@higgs.phys.kyushu-u.ac.jp}}\\%
     {\footnotesize \it%
     Department of Physics, Kyushu University,}\\
     {\footnotesize \it%
      Hakozaki, Higashi-ku, Fukuoka 812-8581, Japan%
      }%
   \\{~}\\
   Joe Sato%
     \footnote{e-mail address: {\tt joe@rc.kyushu-u.ac.jp}}\\%
     {\footnotesize \it%
     Research Center for Higher Education, Kyushu University,}\\
     {\footnotesize \it%
     Ropponmatsu, Chuo-ku, Fukuoka 810-8560, Japan%
     }%
   \\
   and
   \\
   Yoshitaka Kuno%
     \footnote{e-mail address: {\tt kuno@phys.sci.osaka-u.ac.jp}}\\%
     {\footnotesize \it%
     Department of Physics, Osaka University,}\\
     {\footnotesize \it%
     Machikane-yama 1-1, Toyonaka, Osaka 560-0043, Japan%
     }%
}
\date{}

\begin{document}

\maketitle

\begin{abstract}
A possibility to search for T-violation in neutrino oscillation with
the use of muon polarization is studied. The sensitivity to
T-violation is examined with various magnitudes of muon polarization 
as a function of muon energy and long-baseline distances.
\end{abstract}

\section{Introduction}

In this note, we discuss a possibility to discover the CP-violating
effect in the leptonic sector \cite{pioneer} at a neutrino factory.  
Observation of CP-violation would imply the measurement of 
an imaginary part of the couplings in the Lagrangian associated 
with leptons \cite{AKS}.  
There are two ways to study CP violation. They are

\begin{list}{}{}
\item 1. to study a difference between the CP conjugate modes, or
\item 2. to study a difference between the T conjugate modes.
\end{list}

The first method is to observe the difference between the particle and
its anti-particle. In a neutrino factory, it can be studied by
comparing the difference of the appearance event rates between
$\nu_e\rightarrow\nu_\mu$ and $\bar\nu_e\rightarrow\bar\nu_\mu$ 
\cite{BGRW, Cern0103, FHL0105, NuFact, BCR, YP0105, KS}.
However, it has been pointed out that the matter effect would
introduce a sizeable fake CP-odd effect \cite{lipari01, KOS}. 
It is necessary to discriminate
the genuine CP violation effect from the fake matter effect.

The second method is to observe the difference between the transitions
$\nu_e\rightarrow\nu_\mu$ and ${\nu_\mu\rightarrow\nu_e}$.  By the CPT
theorem, in vacuum, observation of T-violation is identical to
observation of CP-violation. Even in a terrestrial oscillation
experiment, since a possible T-odd effect from asymmetric matter
density profile in the earth is very small \cite{takasugi01}, this
T-reversed difference reveals as a very clean signal \cite{AKS, HS} 
on the imaginary phase in the lepton sector \cite{AKS, KS, lipari01, 
takasugi01, HS, rubbia01, lindner01, PW, blondel00, YKT, Andre, 
Yasuda99, 
Barger99, Bil98, T1, KrPe, KP, Sato00}.

Therefore, the comparison of T-reversed oscillation modes would give
clear signature than those of CP-reversed oscillation modes. However,
although it is easy to observe $\nu_{e} \rightarrow \nu_{\mu}$
oscillation, it is known to be difficult to observe $\nu_{\mu}
\rightarrow \nu_{e}$ oscillation at a neutrino factory. The reason is
as follows. At a neutrino factory, $\nu_\mu$ is generated by the decay
of muons as well as $\bar\nu_e$. The former after the oscillation from
$\nu_{\mu} \rightarrow \nu_{e}$, produces $e^{-}$s and the latter from
muon decays produces $e^{+}$s at a detector. It is difficult to
distinguish $e^+$ from $e^-$ after the creation of electromagnetic
shower in most of the detectors being considered at
present. Therefore, it is difficult to identify the oscillation
events from the non-oscillation events which is larger in statistics.

It is critical to examine whether there are any solutions to solve
this problem for the search for T-violation modes useful.  There was
one idea to identify the oscillation events by the use of polarized
muons in the muon storage ring \cite{blondel00}. The principle of the
idea is the following. Supposed that the muons in the muon storage
ring have their spin polarization $P_{\mu^{-}}=-1$, there would be no
$\bar{\nu}_{e}$ and only $\nu_{\mu}$ at the very forward direction
along the muon momentum direction. Observation of $e-$like events at a
detector would directly imply the appearance of $\nu_{\mu} \rightarrow
\nu_{e}$. In reality, the muon spin polarization would not be 100
\%. But, if the spin polarization of the muons in the muon storage
ring can be changed, accordingly the neutrino energy spectrum shapes
can be changed. The changes of the spectrum of neutrino and
anti-neutrino are different. Therefore, in principle we could
discriminate the events associated with neutrinos from those from
anti-neutrino. In turn, the oscillating events and non-oscillating
events can be separated.


\section{Principle}

In this section, the principle to discriminate the oscillating events
by using the muon polarization is presented. The neutrino flux in the
very forward direction of a $\mu^-$ beam with its muon spin
polarization $P_{\mu}$ is given by

\begin{eqnarray}
 \Phi_{\bar\nu_e} &\propto& y^2 (1 - y) \cdot (1 + P_{\mu}),
\label{nu_eFlux}
\\ \Phi_{\nu_\mu} &\propto& y^2 (3 - 2y) + P_{\mu} \cdot y^2 (1 - 2y),
\label{nu_muFlux}
\end{eqnarray}

\noindent where $y\equiv E_\nu/E_\mu$. $E_{\mu}$ is the $\mu^-$ energy
and $E_\nu$ is the neutrino energy.  For $\mu^{+}$,
eqs.(\ref{nu_eFlux}) 
and (\ref{nu_muFlux})
can be given with changing the sign of $P_{\mu}$. From using
eqs.(\ref{nu_eFlux}) 
and (\ref{nu_muFlux}), the non-oscillating event rate of
$\bar\nu_e\rightarrow\bar\nu_e$ ($N_{\bar{\nu}_e \rightarrow
\bar{\nu}_e}$) and the appearance event rate of
$\nu_\mu\rightarrow\nu_e$ ($N_{\nu_\mu\rightarrow\nu_e}$) are given by

\begin{eqnarray}
 N_{\bar\nu_e\rightarrow\bar\nu_e}=( 1+ P_{\mu} ) N_3,
\label{nu_eEvent}
\\
 N_{\nu_\mu\rightarrow\nu_e}=N_1+{P_{\mu}} N_2,
\label{nu_muEvent}
\end{eqnarray}

\noindent respectively. Here $N_1$, $N_2$ and $N_3$ are the event 
numbers
corresponding to each term of neutrino flux.  
They are typically represented by
\begin{eqnarray}
 N_{1} &=& C \frac{E_{\mu}^{2}}{L^{2}}
           \int_{E_{{\rm min}} / E_{\mu}}^{E_{{\rm max}} / E_{\mu}} 
dy 
            2 y^{2} (3-2y) 
            P_{\nu_{\mu} \rightarrow \nu_{e}}(E_{\nu})
            \sigma(E_{\nu}),\\
 N_{2} &=& C \frac{E_{\mu}^{2}}{L^{2}}
           \int_{E_{\rm min}/ E_{\mu}}^{E_{\rm max} / E_{\mu}} dy 
            2 y^{2} (1-2y) 
            P_{\nu_{\mu} \rightarrow \nu_{e}}(E_{\nu})
            \sigma(E_{\nu}),\\ 
 N_{3} &=& C \frac{E_{\mu}^{2}}{L^{2}}
           \int_{E_{\rm min} / E_{\mu}}^{E_{\rm max} / E_{\mu}} dy 
            12 y^{2} (1-y) 
            P_{\bar{\nu}_{e} \rightarrow \bar{\nu}_{e}}(E_{\nu})
            \bar{\sigma}(E_{\nu}),
\end{eqnarray}

\noindent where $P_{\nu_{\mu} \rightarrow \nu_{e}}$, $P_{\bar{\nu}_{e}
\rightarrow \bar{\nu}_{e}}$ are oscillation probabilities, $\sigma$
($\bar{\sigma}$) is the detection rate including the neutrino
(anti-neutrino) cross section and signal efficiency, $E_{\rm min}$
($E_{\rm max}$) is minimum (maximum) neutrino energy observed in this
assumption, $ L $ is the baseline length, and $C$ is the constant
proportionate to the decay muon number and the detector mass.  From
these equations, the rate of $e-$like events, which is the sum of
$N_{\bar{\nu}_e \rightarrow \bar{\nu}_e}$ and
$N_{\nu_\mu\rightarrow\nu_e}$, can be given by

\begin{eqnarray}
 N_{e{\rm -like}}&=&\beta_0+{P_{\mu}}\beta_1,
\label{e-likeEvent}\\
\beta_0&\equiv&N_1+N_3,
\label{defBeta0}\\
\beta_1&\equiv&N_2+N_3.
\label{defBeta1}
\end{eqnarray}

\begin{figure}
    \unitlength=1cm
    \begin{picture}(15,9)
        \includegraphics[width=15cm]{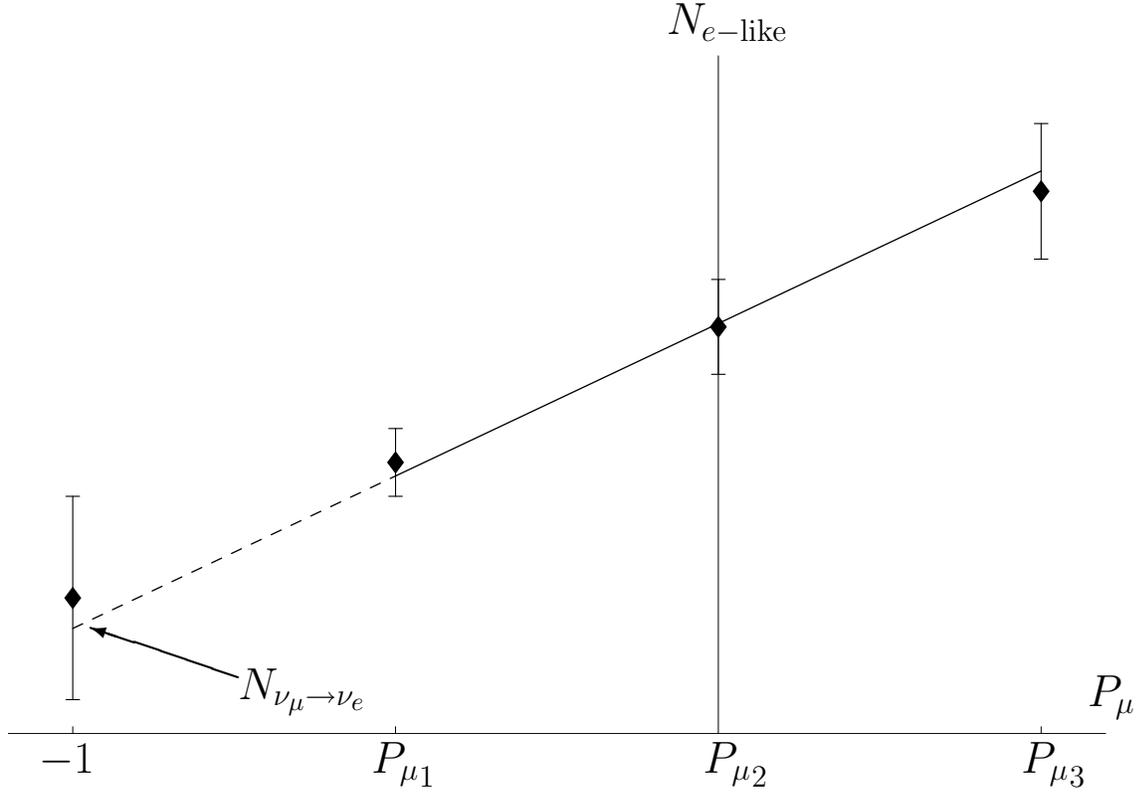}
    \end{picture}
    \caption{Graphical view of the idea. We have observations of 
$e-$like
event with partially polarized muon beams and from them we estimate 
the
event rate at ${P_{\mu}}=-1$.}
    \label{fig:idea}
\end{figure}

\noindent Note that for ${P_{\mu}}=-1$, $N_{e {\rm -like}}$ becomes
equal to $N_{\nu_\mu\rightarrow\nu_e}$ itself from
eq.(\ref{nu_eEvent}).  The estimate of the $e-$like event rate at
${P_{\mu}}=-1$ is essentially the measurement of
$N_{\nu_\mu\rightarrow\nu_e}$. In another words, if $\beta_0$ and
$\beta_1$ can be estimated, the direct observation of
$N_{\nu_\mu\rightarrow\nu_e}$ could be made in principle.  This is
shown in Fig.\ref{fig:idea}.

\section{Procedure}

In this section, the procedure to estimate the appearance event rate,
$N_{\nu_\mu\rightarrow\nu_e}$, at ${P_{\mu}}=-1$ from the measurements
is given. In reality, the muons in the muon storage ring have some
distribution of their polarization. It is assumed that a number of the
muons of their muon polarization of ${P_{\mu}}_i$, $(i=1\cdots n)$ is 
$N_i
(\equiv N_{\mu} / f_{i}$), $(i=1 \cdots n)$. They yield $y_i$ $e-$like
events. $N_{\mu}$ is a total number of the muons in the ring.  For
those, eq.(\ref{e-likeEvent}) can be rewritten by

\begin{eqnarray}
 N_{e {\rm -like}}&=&\beta_0^*+{P_{\mu}}^*\beta_1,
\label{e-likeEvent2}\\
\beta_0^*&\equiv&\beta_0+\bar{P_{\mu}}\beta_1,
\label{defBeta*}\\
{P_{\mu}}^*&\equiv&{P_{\mu}}-\bar{P_{\mu}}=-1-\bar{P_{\mu}},
\label{defAlpha*}\\
\bar{P_{\mu}}&\equiv&\frac{1}{F}\sum_i\frac{{P_{\mu}}_i}{f_i},
\label{defBarAlpha}\\
F&\equiv&\sum_i\frac{1}{f_i}.
\label{defF}
\end{eqnarray}

The expectation values of $y_i$, $\bar y_i$ is given by%
\footnote{In the following, we denote $\beta_0^*$ by $\beta_0$.}

\begin{eqnarray}
 f_i \bar y_i=\beta_0+{P_{\mu}}^*_i\beta_1.
\label{binExpectation}
\end{eqnarray}

It is assumed that $y_i$'s are large enough so that it would follow
the normal distribution of ${\rm N}(\bar y_i,\bar y_i)$.  In this 
case, the
likelihood function to estimate $\beta_0$ and $\beta_1$ is given by%
\footnote{Exactly speaking, the denominator of eq.(\ref{likelyfood})
$f_i^2 \bar y_i$. However, for simplicity, under the assumption that
$y_i$'s are large, it can be replaced with $f_i^2 y_i$. This is
justified since the non-oscillating event rates
$N_{\bar\nu_e\rightarrow\bar\nu_e}$ at any polarization are large.
(It does not depend on whether there are enough appearance event rate
$N_{\nu_\mu\rightarrow\nu_e}$ as long as the polarization is not very
close to $-1$.) } \footnote{The likelihood function with Poisson
distribution can be also treated. By this, the same function as
eq.(\ref{likelyfood}) can be obtained under the assumption that
$y_i$'s are large enough. }

\begin{eqnarray}
-\log L = \sum_i\frac{\left\{f_i 
y_i-(\beta_0+{P_{\mu}}^*_i\beta_1)\right\}^2}
{2 f_i^2 y_i}.
\label{likelyfood}
\end{eqnarray}

\noindent From this likelihood function, the equations for the
estimates of $\beta_0$ and $\beta_1$ can be give by

\begin{eqnarray}
0 = -\frac{\partial\log L}{\partial \beta_0}&=&
\beta_0\sum_i\frac{1}{f_i^2 y_i}+
\beta_1\sum_i\frac{{P_{\mu}}_i^*}{f_i^2 y_i}
-\sum_{i} \frac{1}{f_i},
\label{condBeta0}
\\
0 = -\frac{\partial\log L}{\partial \beta_1}&=&
\beta_0\sum_i\frac{{P_{\mu}}^*_i}{f_i^2 y_i}+
\beta_1\sum_i\frac{({P_{\mu}}_i^*)^2}{f_i^2 y_i}
-\sum_{i} \frac{{P_{\mu}}^*_i}{f_i}
\label{condBeta1}\\
&=&
\beta_0\sum_i\frac{{P_{\mu}}^*_i}{f_i^2 y_i}+
\beta_1\sum_i\frac{({P_{\mu}}_i^*)^2}{f_i^2 y_i}.
\nonumber
\end{eqnarray}

Here the definitions of ${P_{\mu}}^*$ by eq.(\ref{defAlpha*}), and of
$\bar{P_{\mu}}$ by eq.(\ref{defBarAlpha}) are used. Also the last term
in the first line of eq.(\ref{condBeta1}) is removed.  Using these
eqs.(\ref{condBeta0}) and (\ref{condBeta1}), $\beta_0$ and $\beta_1$
in terms of $y_i$ can be estimated as follows:

\begin{eqnarray}
 \beta_0&=&\frac{F}{D}\sum_i\frac{({P_{\mu}}_i^*)^2}{f_i^2 y_i} \quad
{\rm and}
\label{estimateBeta0}\\
 \beta_1&=&-\frac{F}{D}\sum_i\frac{{P_{\mu}}_i^*}{f_i^2 y_i},
\label{estimateBeta1}
\end{eqnarray}

\noindent where

\begin{eqnarray}
 D\equiv \left(\sum_i\frac{1}{f_i^2 y_i}\right)
\left(\sum_i\frac{({P_{\mu}}_i^*)^2}{f_i^2 y_i}\right)
-\left(\sum_i\frac{{P_{\mu}}_i^*}{f_i^2 y_i}\right)^2.
\label{defD}
\end{eqnarray}

Thus, $N_{\nu_\mu\rightarrow\nu_e}$ at ${P_{\mu}}=-1$ can be estimated
using eqs.(\ref{estimateBeta0}) and (\ref{estimateBeta1}) as
\begin{eqnarray}
 N_{\nu_\mu\rightarrow\nu_e}=\beta_0+(-1-\bar{P_{\mu}})\beta_1.
\label{estimateNbar}
\end{eqnarray}

The statistical error on this estimated value is now evaluated.
Formally, the variance of the estimate is given by

\begin{eqnarray}
 V(N_{\nu_\mu\rightarrow\nu_e})=V(\beta_0+(-1-\bar{P_{\mu}})\beta_1)
=V(\beta_0)+(-1-\bar{P_{\mu}})^2
V(\beta_1)+2(-1-\bar{P_{\mu}})V(\beta_0,\beta_1),
\end{eqnarray}

\noindent where $V(\beta)$ is the variance of $\beta$, and
$V(\beta_0,\beta_1)$ is the covariance of $\beta_0,\beta_1$.

It is impossible to evaluate the variance of
$N_{\nu_\mu\rightarrow\nu_e}$ using eqs.(\ref{estimateBeta0}) and
(\ref{estimateBeta1}) directly. Therefore, it is approximated by the
inverse matrix of the expectation value of Fisher's information.%
\footnote{Strictly speaking, the inverse matrix of the expectation
value of Fisher's information gives the lower limit of the
corresponding variance. However if the likelihood function is
constructed well then it coincide with the variance and hence we use
this approximation.  We also approximate the expectation value of the
Fisher's information by eq.(\ref{Fishers}) since it is assumed that
$y_i$'s are large number enough.}  Fisher's information matrix is 
given
by

\begin{eqnarray}
- \frac{\partial^2}{\partial \beta_i\partial\beta_j} \log L =
\left(
\begin{tabular}{cc}
$\displaystyle
\sum_i\frac{1}{f_i^2 y_i}$&
$\displaystyle
\sum_i\frac{{P_{\mu}}_i^*}{f_i^2 y_i}$\cr
$\displaystyle
\sum_i\frac{{P_{\mu}}_i^*}{f_i^2 y_i}$&
$\displaystyle
\sum_i\frac{({P_{\mu}}_i^*)^2}{f_i^2 y_i}
$\end{tabular}
\right),
\label{Fishers}
\end{eqnarray}
and hence the variance matrix is given by
\begin{eqnarray}
\frac{1}{D}
\left(
\begin{tabular}{cc}
$\displaystyle
\sum_i\frac{({P_{\mu}}_i^*)^2}{f_i^2 y_i}$&
$\displaystyle
-\sum_i\frac{{P_{\mu}}_i^*}{f_i^2 y_i}$\cr
$\displaystyle
-\sum_i\frac{{P_{\mu}}_i^*}{f_i^2 y_i}$&
$\displaystyle
\sum_i\frac{1}{f_i^2 y_i}
$\end{tabular}
\right).
\end{eqnarray}
Thus, the estimate of $N_{\nu_\mu\rightarrow\nu_e}$ has its
variance of
\begin{eqnarray}
 V(N_{\nu_\mu\rightarrow\nu_e})&=&\frac{1}{D}
\left\{\sum_i\frac{({P_{\mu}}_i^*)^2}{f_i^2 y_i}
+(-1-\bar{P_{\mu}})^2 \sum_i\frac{1}{f_i^2 y_i}
-2(-1-\bar{P_{\mu}})\sum_i\frac{{P_{\mu}}_i^*}{f_i^2 y_i}
\right\}
\\
&=&\frac{1}{F}\left[2 N_{\nu_\mu\rightarrow\nu_e}
+N_{\bar{P_{\mu}}}\left\{(-1-\bar{P_{\mu}})^2\frac{\sum_i 1/f_i^2 y_i}
{\sum_i ({P_{\mu}}_i^*)^2/f_i^2 y_i}-1
\right\}
\right].
\label{estimateVbar}
\end{eqnarray}

Here $N_{\bar{P_{\mu}}}$ is the estimate of the event rate at
${P_{\mu}}=\bar{P_{\mu}}$ when $N_{\mu^-}$ $\mu^-$s are used.

\section{Numerical Analysis}
\label{sec:Numerical}

In the analytical treatment shown above, several assumptions are 
made. For instance, it is assumed that the likelihood function of the 
$\beta$'s is given by eq.(\ref{likelyfood}), and the variance of the 
estimate is given by the inverse of the Fisher's information matrix.  

In this section, numerical calculations is given. What we have to 
calculate numerically is  the extrapolated value at ${P_{\mu}}=-1$
which indeed coincides with the theoretical expectation value $\bar
y|_{{P_{\mu}}=-1}$ and the variance of it which is actually given by
eq.(\ref{estimateVbar}).%

The algorithm for this numerical analysis is as follows:

\begin{enumerate}
\item [step 0.] Select several values of the polarization 
${P_{\mu}}_i$
for the measurements,
\item[step 1.] Fix all the parameters, such as the theoretical 
parameters
(e.g. mixing angle), a muon energy and so on, and then calculate the
expectation of event rates in each polarization values ($\bar y_i$).
\item[step 2.] Generate the ``actual'' events of $y_i$ according to 
Poisson
distribution with its expectation value $\bar y_i$
\item[step 3.] Substitute these event sets into
eqs.(\ref{estimateBeta0}), (\ref{estimateBeta1}), and find the
extrapolated value at ${P_{\mu}}=-1$ from eq.(\ref{estimateNbar}) by
using these $\beta$'s.
\item[step 4.] Iterate the step 2 and step 3 in several times to find
the type of distribution of the extrapolated value.
\end{enumerate}

\noindent This virtual measurements for various parameters were
made. It is found that the extrapolated value follows the normal
distribution with its mean value $\bar y|_{{P_{\mu}}=-1}$ and its
variance given by eq.(\ref{estimateVbar}).

\section{Sensitivity for T Violation}

\begin{figure}
    \unitlength=1cm
    \begin{picture}(16,16)
        \includegraphics[width=16cm]{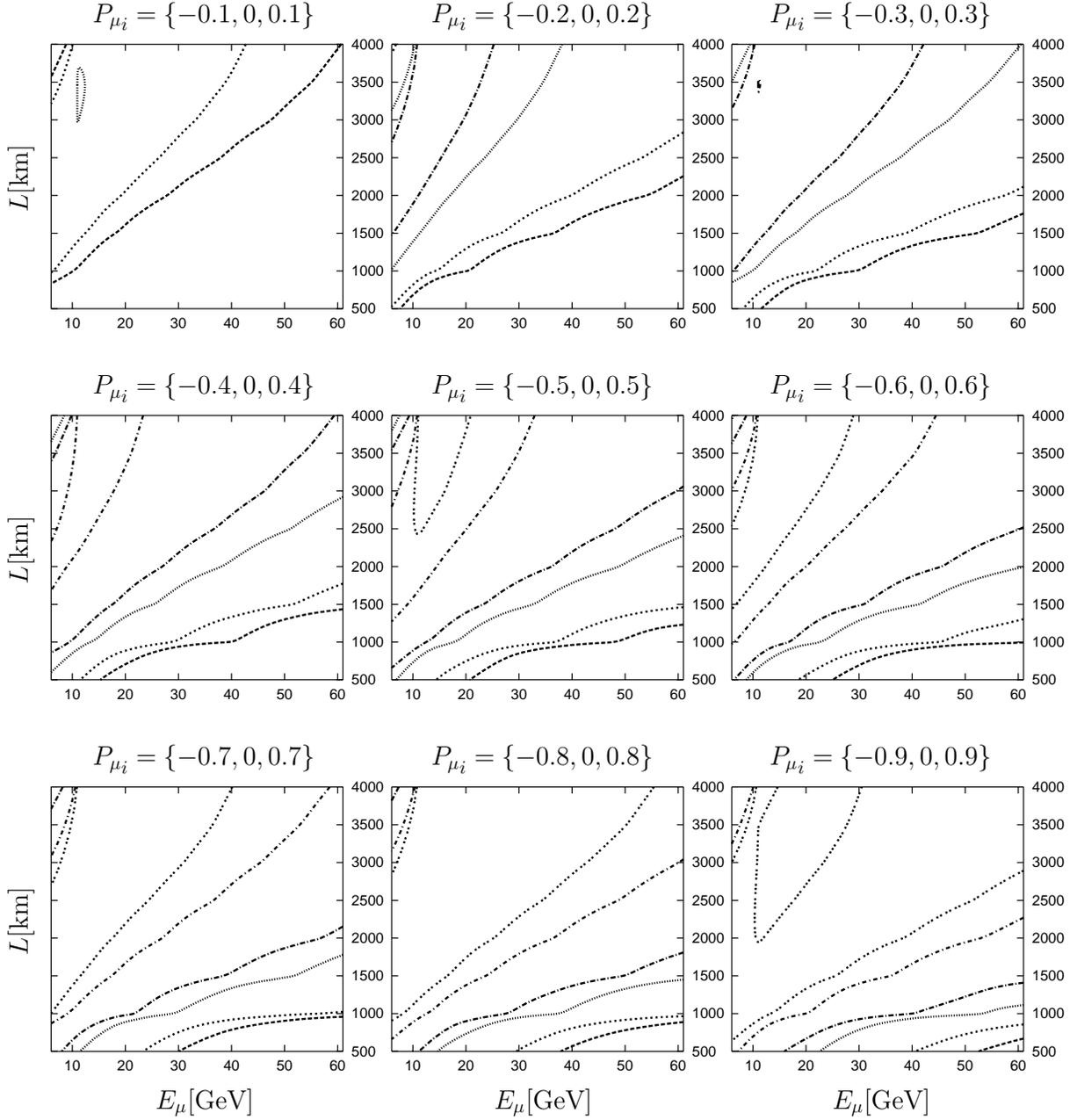}
    \end{picture}
    \caption{Required data sizes in the unit of $10^{21}$ muon decays
             $\times$ 100-kt detector to obtain a sensitivity of $90\%$
             confidence level are plotted as a function of muon energy
             and baseline length for various sets of muon
             polarizations ${P_{\mu}}_{i}$. 
             The meaning of each line is explained in 
             Fig.\ref{fig:alphaplot-dm5e-05}.
             For example, the contour
             with the label of 0.5 implies $5 \times 10^{20}$ muon
             decays with a 100-kt detector is used. For each case,
             three measurements with different ${P_{\mu}}_{i}$ values
             are assumed. These three values of the muon polarization
             are shown for each plot. If the muon polarization is
             high, the necessary data size is smaller.}
\label{fig:alphaplot}
\end{figure}

In this section, the sensitivity to T-violation using the present
procedure is discussed to find the statistics to see the T-violation
effect.

\begin{eqnarray}
 \chi^2(n)&=& \sum_{i}^{n} 
              \frac{(N_{0 i} \bar{N}_{i} -\bar{N}_{0 i} N_{i})^2}
                   {N_{0 i}^2 V_{i} + \bar N_{0 i}^2 N_{i}},
\label{chi2}
\end{eqnarray}
where
\begin{eqnarray}
N&:&N_{\nu_e\rightarrow\nu_\mu} ,\nonumber\\
\bar N&:&N_{\nu_\mu\rightarrow\nu_e} ,\nonumber
\end{eqnarray}

\noindent and the subscript 0 indicates the estimate of the event rate
with CP violating phase $\delta=\delta_0 \equiv \{0,\pi\}$ and
subscript $i$ represents $i$th energy bin. Note that in the
denominator of $\chi^2$ there is $V$.  Although $V$ is given by
eq.(\ref{estimateVbar}) theoretically, it is calculated by using the
numerical method in section \ref{sec:Numerical}. To confirm that T
violation can be observed at $90 \%$ confidence level, $\chi^{2}(n) >
\chi^{2}_{90 \%} (n)$ is required.  Suppose that three measurements
can be adopted with different muon polarization values ($\pmu_{i} =
\{ -\pmu, 0, \pmu \}$),
the sensitivity of $90 \%$ confidence level can be calculated. It is
shown in Fig.\ref{fig:alphaplot}. We chose the most sensitive method
among $n=$ 1, 3, 5 in each ($E_{\mu}$, $L$) region. The usage of
statistics is referred in Ref.\cite{KOS} in detail.

\begin{figure}
    \unitlength=1cm
    \begin{picture}(15,15)
        \includegraphics[width=15cm]{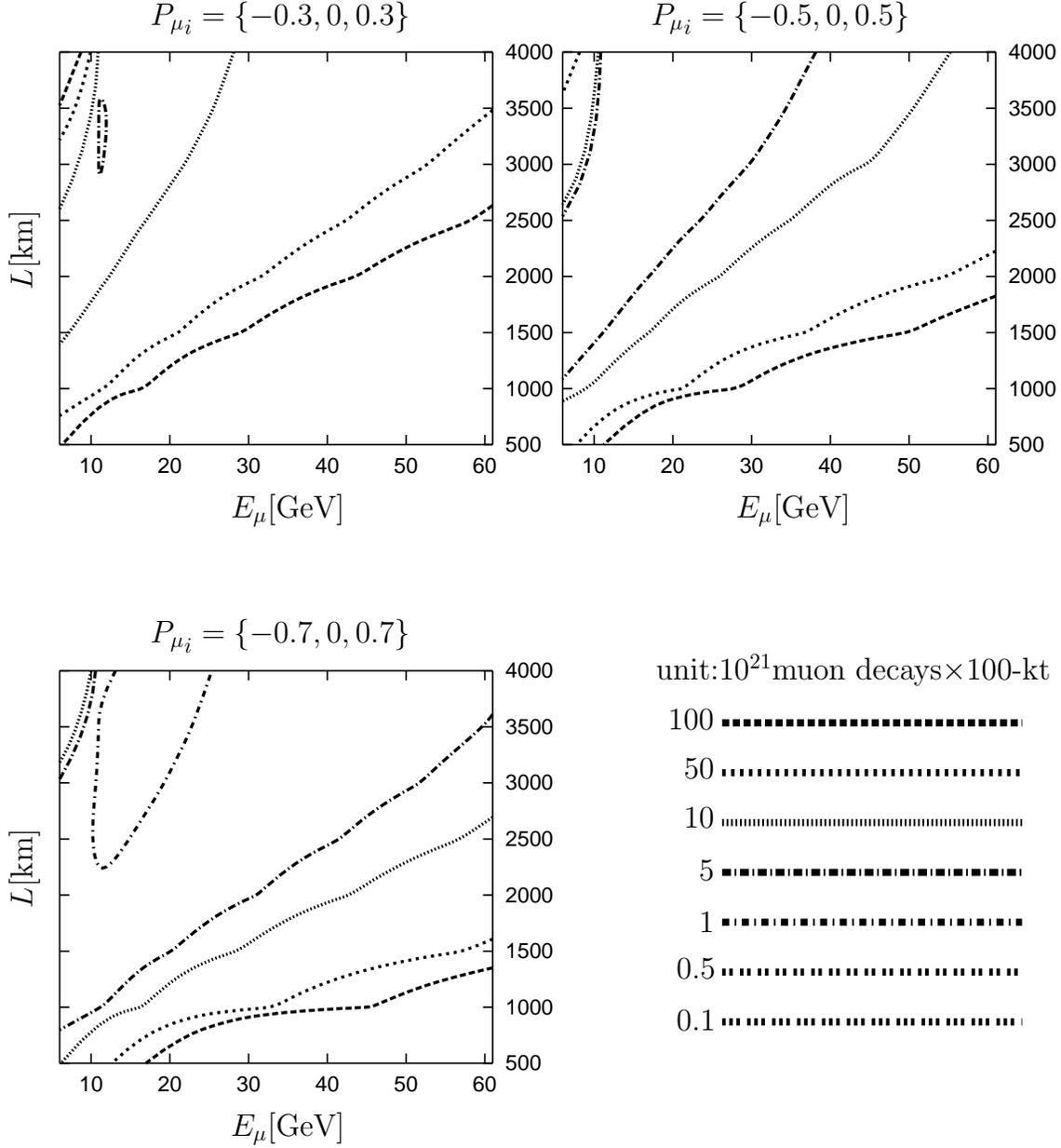}
    \end{picture}
    \caption{Same as Fig.\ref{fig:alphaplot}, but here 
             $\delta m_{21}^{2}=5\times 10^{-5}{\rm eV^{2}}$}
\label{fig:alphaplot-dm5e-05}
\end{figure}

Here, it is discussed how many neutrinos from muon decays should be
needed to observe the T-violation effect. First of all, for
simplicity, $N_i=N_{\mu^+}$ is assumed. Here, the following
theoretical parameters are used; $\sin\theta_{12}=0.5$,
$\sin\theta_{23}=1/\sqrt{2}$, $\sin\theta_{13}=0.1$, $\delta
m^2_{31}=3\times 10^{-3} {\rm eV}^2$, $\delta m^2_{21}=10^{-4}$eV$^2$,
$\delta=\pi/2$. We assume that each parameter including matter density
has 10$\%$ ambiguity.  The sensitivity is almost proportional to

\begin{eqnarray}
\left(\sin\delta \sin 2\theta_{12} \sin 2\theta_{23}
 \cos\theta_{13}
\frac{ \delta m_{21}^2}{ \delta m_{31}^2} \right)^2.
\end{eqnarray}
Therefore, the sensitivity becomes 4 times worse 
when $\delta m_{21}^{2}=5 \times 10^{-5}\rm{eV}^{2}$ 
(see Fig.\ref{fig:alphaplot-dm5e-05}). 

In Fig.\ref{fig:alphaplot}, it is found that at worst $P_{\mu}=\pm0.3$
should be needed to observe T-violation effect with a total of
$10^{21}$ muon decays and a 100-kt detector. On the contrary to the
sensitivity of CP-violation effect usually discussed, the sensitivity
will not change drastically by including other sources of errors,
since the $V$ will convey the largest error among the other
uncertainties.

In the remaining of this section, we see what we can learn from the
theoretical analysis.

\subsection{How many $\mu^+$ and $\mu^-$ decays are needed ?}

In this subsection, the most efficient ratio, $f_i$, between the
numbers of $\mu^+$ $(N_{\mu^+})$ and $\mu^-$ $(N_{\mu^-})$ to obtain
the best sensitivity to CP violation is estimated for a fixed
distribution of $ N_{\mu^-}$ at each polarization ${P_{\mu}}_i$. To do
this, the $\chi^2$ can be rewritten to factor out the numbers of
events as follows,

\begin{eqnarray}
N&\rightarrow&N_{\mu^+} N ,\nonumber\\
\bar N&\rightarrow& N_{\mu^-} \bar N,\nonumber\\
V &\rightarrow& N_{\mu^-} V,\nonumber
\end{eqnarray}
\begin{eqnarray}
 \chi^2&\rightarrow&\frac{(N_0 \bar N-\bar N_0 N)^2}{N_0^2 V/
N_{\mu^-} + \bar N_0^2 N/N_{\mu^+}}.
\label{chi2v2}
\end{eqnarray}

\noindent If the total numbers of muon is given, then

\begin{eqnarray}
 \sum_{i} N_i + N_{\mu^+}=
\sum_i\frac{N_{\mu^-}}{f_i}+N_{\mu^+}=F N_{\mu^-}+N_{\mu^+}=N
\end{eqnarray}

\noindent is the given number of the total events. Here, $F$ is given
by eq.(\ref{defF}).  Under these conditions, the most efficient ratio
of the numbers of $\mu^+(N_{\mu^+})$ and $\mu^-( N_{\mu^-})$ is given
by

\begin{eqnarray}
 \frac{N_{\mu^+}}{\sum_{i} N_i}=\frac{\bar 
N_0}{N_0}\sqrt{\frac{N}{A}},
\end{eqnarray}

\noindent where $A$ is given by eq.(\ref{estimateVbar}) $\times F$.

Since the oscillation probability is at most a few \%, while
$N_{\bar{P_{\mu}}}$ in eq.(\ref{estimateVbar}) is essentially given
by the non-oscillation event, from the fact that $N/N_{\bar{P_{\mu}}}
\sim O(10^{-2})$ and that $\sum_i [1/(f_i^2 y_i)]/\sum_i
[({P_{\mu}}_i^*)^2/(f_i^2 y_i)]$ in eq.(\ref{estimateVbar}) is much
larger than a unity in a realistic case, it is found that a much
larger number of $\mu^-$ is needed. Therefore by tuning the ratio of
the numbers of $\mu^+$ and $\mu^-$, a higher sensitivity to
T-violation effect can be obtained by several tens \%.

\subsection{Dependence of Muon Polarization Distribution}

The measurements of $N_{i}$ events with muon polarization
${P_{\mu}}_{i}$ are usually made. In this subsection, the optimization
of $N_{i}$s ($i=1 \cdots n$) for each given ${P_{\mu}}_{i}$ to obtain 
the best
sensitivity is studied.  This is equivalent to examine the best muon
polarization distribution.


\begin{figure}
    \unitlength=1cm
    \begin{picture}(15,9)
        \includegraphics[width=15cm]{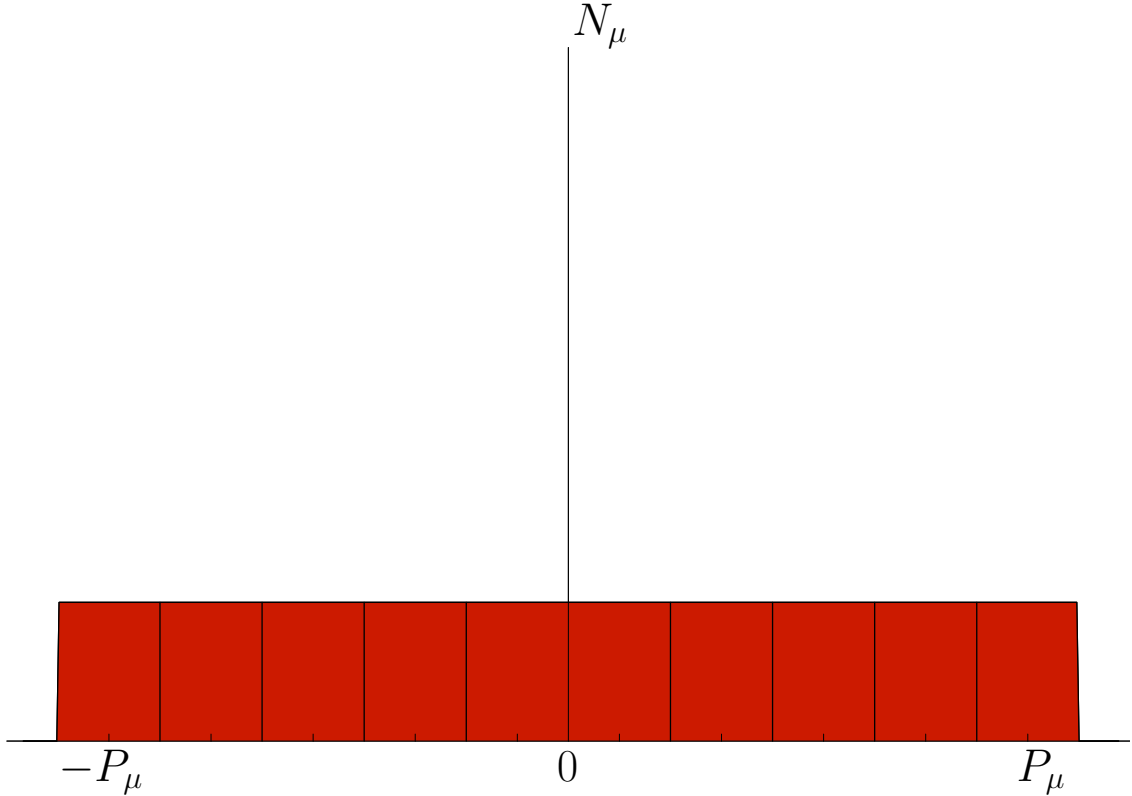}
    \end{picture}
    \caption{Uniform Distribution of polarization of muon, being
    distributed from $-{P_{\mu}}$ to $+{P_{\mu}}$ uniformly.}
    \label{fig:continuousP}
\end{figure}

For the most simple case to consider is a uniform distribution of
$N_{i}$ from $-{P_{\mu}}$ to $+{P_{\mu}}$ as shown in
Fig.\ref{fig:continuousP}.

\begin{figure}
    \unitlength=1cm
    \begin{picture}(16,10)
        \includegraphics[width=15cm]{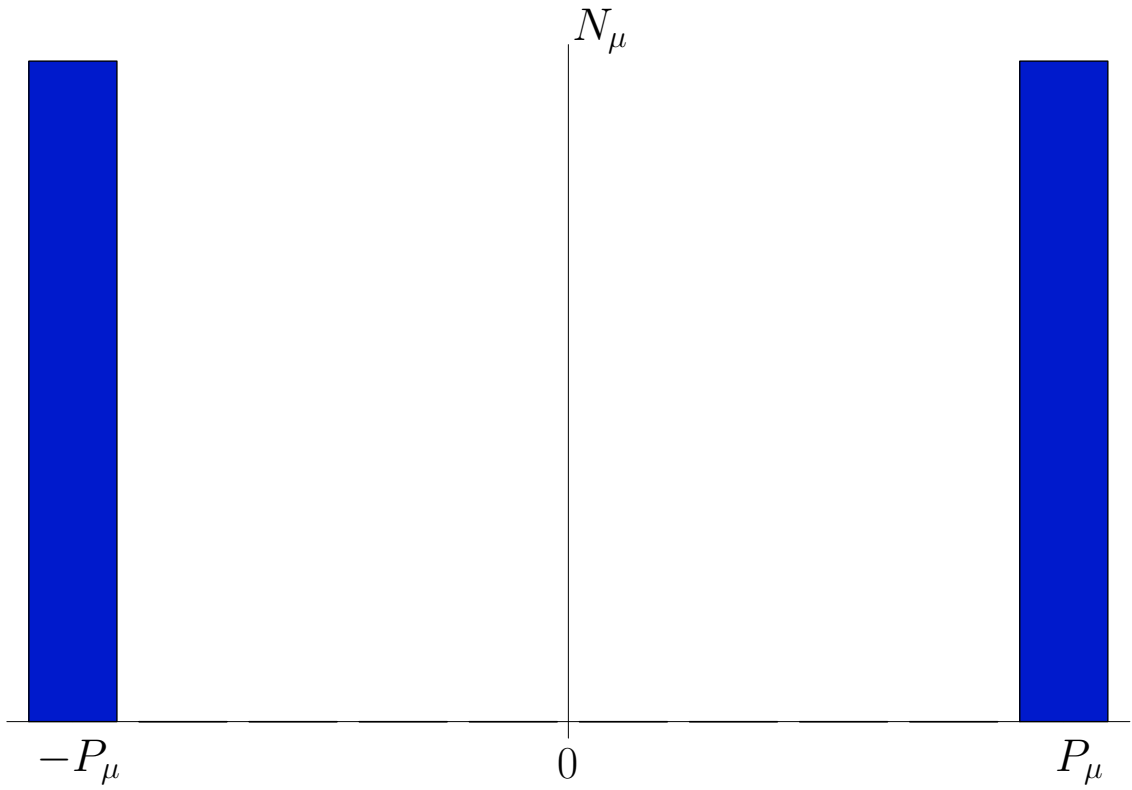}
    \end{picture}
    \caption{Distribution of the polarization of $\mu$.  There are 
only
    muons with their polarization around $-{P_{\mu}}$ and 
$+{P_{\mu}}$.}
    \label{fig:distinctP}
\end{figure}

Let's consider two cases. One is a uniform distribution of $N_{i}$ 
from
$-{P_{\mu}}$ to $+{P_{\mu}}$ as shown in Fig.\ref{fig:continuousP}.
The other is discrete measurement at $\pm P_{\mu}$ as seen in
Fig.\ref{fig:distinctP}.

It is assumed that the same numbers of muon decays are used for the
two different distributions of Fig.\ref{fig:continuousP} and
Fig.\ref{fig:distinctP}. In this case, the variances
(\ref{estimateVbar}) of the extrapolated appearance numbers are
different, depending on the distribution. In case of
Fig.\ref{fig:continuousP},%
\footnote{Here we assume that we do not have very large polarization.
It means that $y_i$'s are almost constant.}

\begin{eqnarray}
 \frac{1}{F} \frac{\sum_i 1/f_i^2 y_i}
{\sum_i ({P_{\mu}}_i^*)^2/f_i^2 y_i}-1 \propto 
\frac{3}{{P_{\mu}}^2}-1,
\end{eqnarray}

\noindent while in case of Fig.\ref{fig:distinctP},

\begin{eqnarray}
 \frac{1}{F} \frac{\sum_i 1/f_i^2 y_i}
{\sum_i ({P_{\mu}}_i^*)^2/f_i^2 y_i}-1 \propto 
\frac{1}{{P_{\mu}}^2}-1.
\end{eqnarray}

\noindent Therefore, a three times higher sensitivity can be obtained
for the distribution of Fig.\ref{fig:distinctP}.
In reality, there will be a loss of muon decays to create the
distribution in Fig.\ref{fig:distinctP}. If this reduction factor of
the number of events is defined to be M (namely only (1/M) events can
be used), the sensitivity is given by

\begin{eqnarray}
\frac{1}{F} \frac{\sum_i 1/f_i^2 y_i}
{\sum_i ({P_{\mu}}_i^*)^2/f_i^2 y_i}-1 \propto 
\frac{M}{{P_{\mu}}^2}-1.
\end{eqnarray}

Then the sensitivity is better by $3/M$. To discuss further, it is
needed to have realistic muon polarization distribution. However, 
once 
it is obtained, the discussion to study a sensitivity is quite
straightforward.

\section{Summary and Discussion}

In this note, we have examined a possibility to study T-violation with
the use of muon polarization. The physics motivation to search for 
T-violation is to observe direct CP violation arising from the 
imaginary phase in the neutrino mixing matrix. The advantage of 
search for T-violation is to have a negligible small matter effect. 

In the present study, the sensitivity to T-violation has been studied
as a function of muon energy and long baseline length for various set
of muon polarizations.  In each set, it is assumed that at minimum
three measurements with different values of muon polarization are used,
for instance, ${P_{\mu}}_{i}$ = $\pm P_{\mu}$ and $0$ are used in the 
present study, where $P_{\mu}$ is the magnitude of the muon
polarization.  Needless to say, it is shown that the muon polarization
achievable is larger, the sensitivity is better.  From the present
study, it is found that $|P_{\mu}|> 0.3$ is necessary when it is assumed
that we can get a total of $10^{21}$ muon decays and a 100-kt detector.

The magnitude of muon polarization is being studied in the
accelerator study. It is not easy to have large muon polarization in 
a very high intensity. It would be needed that the magnitude of muon 
polarization and a number of muons have to be traded off. However, if 
search for T violation can be made with realistic magnitude of muon 
polarization, it would provide unique way to study the CP phase in 
the MNS leptonic mixing matrix.

\subsection*{Acknowledgments}

The work of J.S. is supported  in part by a Grant-in-Aid for 
Scientific
Research of the Ministry of Education, Science and Culture,
\#12047221, \#12740157.


\end{document}